\documentclass[12pt]{iopart}  

\RequirePackage{graphicx}
\RequirePackage[colorlinks,citecolor=blue,urlcolor=blue,linkcolor=blue]{hyperref}
\usepackage{academicons}
\usepackage{xcolor}
\newcommand{\orcid}[1]{\href{https://orcid.org/#1}{\textcolor[HTML]{A6CE39}{\aiOrcid}}}
\usepackage{siunitx}
\begin{document}
\title{Investigating 3D printed Cartesian divers
}



\author{Jonas Bley$^1$, Antony Pietz$^2$, Angela Fösel$^3$, Michael Schmiedeberg$^4$, Stefan Heusler$^5$, Alexander Pusch$^6$}

\address{$^1$ Westfälische Wilhelms-Universität Münster, 48149 Münster, Germany}
\address{$^2$ Westfälische Wilhelms-Universität Münster, 48149 Münster, Germany}
\address{$^3$ Didaktik der Physik, Department Physik, Friedrich-Alexander-Universität Erlangen–Nürnberg, Staudtstr. 7, 91058 Erlangen, Germany; e-mail: angela.foesel@fau.de}
\address{$^4$ Institut f\"ur Theoretische Physik 1, Friedrich-Alexander Universität Erlangen-Nürnberg, Staudtstr. 7, 91058 Erlangen, Germany, e-mail: michael.schmiedeberg@fau.de}
\address{$^5$ Institut für Didaktik der Physik, Westfälische Wilhelms-Universität Münster, Wilhelm-Klemm-Str. 10, 48149 Münster, Germany; e-mail: stefan.heusler@wwu.de}
\address{$^6$ Institut für Didaktik der Physik, Westfälische Wilhelms-Universität Münster, Wilhelm-Klemm-Str. 10, 48149 Münster, Germany; e-mail: alexander.pusch@wwu.de}

\begin{abstract}
Despite the difficult circumstances due to the COVID-19 pandemics, physics students can tackle interesting questions that are part of physics competitions as the German Physicists' Tournament (GPT) 2020. Due to the COVID-19 pandemics in 2020, many competitions such as the GPT are held online. Furthermore, the usual options of equipment offered by the supervising university institutions could not be used by the students. The problems of the GPT 2020 therefore had to be chosen in such a way that they could be examined at home using simple means. One of these supposedly simple but profound experiments - the Cartesian divers - is described in this article. By using 3D printing, the relevant variables could be varied in a controlled manner and the theoretical model for Cartesian divers could be examined experimentally.
\end{abstract}

\maketitle

\section{Introduction}
\label{intro}

Cartesian divers have fascinated scientists for a long time - some reports date back to the 17th century \cite{Magiotti} - and are also well known as a toy for children. As they can be used to demonstrate Archimedes' principle and how the buoyancy can change due to the compression of the gas in the divers, Cartesian divers are widely employed for physics education. However, a quantitative analysis of the problem with simple methods is quite difficult, as a systematic variation of the volume of the diver is needed. In this article, we want to show how 3D printing can be used to obtain cheap divers that can be used for well-controlled experiments.

The problem of the cartesian diver that is addressed in this article was listed as problem 11 titled "\emph{Investigating 'Cartesian Devils'}" in the German Physicists' Tournament (GPT) 2020 and reads \cite{gpt-link}:
"\emph{When demonstrating the behavior of a 'Cartesian Devil'/ 'Cartesian diver', that experiment requires a large plastic bottle and a 'diver': either the famous hand blown glass toy from Thuringian forest or simply a small, rigid tube, open at one end. For example, tubes with baking aroma inside are well suited for making a diver. When the Cartesian Devil, filled with water and air, is put in the plastic bottle, filled with water, and when the bottle is pressed, the devil sinks down. By modifying the setup, it is also possible to let it rise. How is this possible? Investigate different possibilities of suitable modifications in order to let the devil rise and discuss the physics behind.}"

Note that the problem of a buoyancy that changes upon compression is not only interesting for educational purposes but is even a topic of recent research in the field of emulsions or colloidal suspensions. For example, the segregation process in mixtures of ultra-soft colloids might lead to complex stacking phases where some particles gather both on the top and on the bottom of a suspension due to the different compression resulting from the pressure gradient in a gravitational field \cite{heras1,kohl,kim,heras2}. Furthermore, extensions of the Archimedes' principle to non-equilibrium systems like shaken granular materials have been widely studied, see e.g. \cite{granular1,granular2,granular3}.

The article is structured as follows: In section \ref{sec:GPT} we give some information related to the Online GPT 2020. In section \ref{sec:didactics} we discuss the importance and potential of the problem and online competitions in physics education. The theoretical background, the experimental realization, and the results of our work on Cartesian divers are presented in section \ref{sec:divers}. Finally, we conclude in section \ref{sec:conclusions}.

\section{Online German Physicist Tournament (GPT) 2020}
\label{sec:GPT}

Founded in 2009, the International Physicists' Tournament (IPT, \cite{IPT}) is a physics competition where students in teams have to work on open questions. At the actual tournament the participants then have to present and defend their approaches in so-called physics fights. In each round of such a fight one team (termed "reporter") presents its solution on a problem, one team (the "opponent") should criticise the approach and propose improvements, while a third team (the "reviewer") collects all arguments of the other teams and moderates the final discussion. The performance of the teams is graded by a jury that usually consists of three to seven experienced physicists. For more details on the concept of IPT (and GPT) see also \cite{IPT_lit}.

The German Physicists' Tournament (GPT) is an annual competition for students in Germany following the rules of the IPT. The winning team of the GPT can usually qualify to represent Germany at the IPT. Furthermore, the problems for a GPT usually are the problems of the next IPT. Due to the COVID-19 pandemics, the IPT 2020 that originally had been scheduled to take part in Warsaw in April 2020 had to be postponed. As a consequence, a delay was expected for the subsequent IPT 2021 such that it became obvious that there will be no regular GPT in 2020. As a replacement, the network that organizes the GPT decided to organize an additional online competition, the Online-GPT 2020. While this online event could not be a qualifying tournament for an IPT and therefore deals with its own problems, it should occur in a way that students who due to the pandemics most likely can not use extensive lab equipment nevertheless are able to work on the problems. As a result, the problem set of the Online GPT 2020 \cite{gpt-link} consists of tasks that usually can be covered at home. For example, one problem was to study patterns that occur when the students bake a zebra cake. Other problems asked to analyse the sound of a water cattle or the sound of rain drops falling on a sheet of metal, which in principal can be done with a smartphone. Further problems dealt with rotating basketballs, caustics of reflected sunlight, of crystals formed by balls floating on a water surface. The complete list of problems can be found online at \cite{gpt-link}.

The organization of an online competition that is usually characterized by personal interactions between the three teams that take part in a physics fight was another challenge. The Online-GPT 2020 took place in December 2020 via the video conference platform \emph{Zoom} \cite{didactics3}. While for the IPT 2020 that finally took place online in September 2020, some rules were adjusted, e.g., except for the final the role of the reviewer was skipped, the Online-GPT 2020 almost completely complied with the regular rules. For clarity, all involved persons had to prepend their name with their role, e.g., "Jury", "Chair", etc., or in case of the participants with the abbreviation of their team. The clock was visible to everyone in an additional video feed.

Obviously, there are some lessons that were learned during the tournament concerning what might be improved in future online competitions. For example, the different stages of a physics fight, e.g., the preparation of the reporter, the report, questions to the reporter, preparation of the opponent, etc., should be more visible to all participants, maybe in an additional video feed or on an additional website that can display the sequence of these stages in real time and in a synchronized way. Furthermore, a timeout for technical reasons probably will be introduced in future online competition in order for participants to deal with technical issues, e.g., if they realize that the slides are shown in a wrong way while they share their screen.

Note that the GPT 2021 as well as the IPT 2021 both will be online again. Furthermore, based on the experiences with the online competition, it has been proposed that at least a part of the GPT might always be organized as an online tournament. Though the personal interactions are limited in online event, an online competition enables more students to take part that usually might not have participated due to the costs or due to long travelling times. The final decision about how online physics fights might be combined with an in-person tournament of the GPT is still pending.

\section{Didactic and pedagogical view of online tournaments}
\label{sec:didactics}

\subsection{The time before COVID-19 pandemic}\label{sec:beforeCOVID}
Usually, science competitions and especially science \emph{tournaments} are carried out with traditional placed-based methods. That means, the organisers of a tournament define a venue, and all participants as well as the organisers and jurors come to that venue and take place at the tournament.

Before COVID-19 pandemic, for the realisation of physics tournaments like \emph{International Young Physicists' Tournament} IYPT (a physics tournament for students at school) or \emph{International Physicists' Tournament} IPT (a tournament for students at university), the subject "online format instead of physical presence" never came up.
Same was true for almost all non-tournament competitions, e.g. for German \emph{Jugend forscht} \cite{didactics1} or for \emph{Interational Science Olympics} \cite{didactics2} like \emph{IJSO}, \emph{EUSO} and \emph{IPhO}.

\subsection{Impact of COVID-19 on educational systems}\label{sec:educationalCOVID}
The pandemic has severely impacted educational systems globally. Most governments have temporarily closed educational institutions, with many switching to online education. Tournaments have been effected, too: The impact of COVID-19 pandemic didn't allow the physical presence of students, organisers and jurors at a venue any more. Hence tournaments could be deleted without replacement or new ideas, ideas for realisations of tournaments using online formats have to be designed. COVID-19 pandemic triggered the birth of online tournaments.

\subsection{Concept of online tournaments}\label{sec:conceps}
The concept of most of the online tournaments carried out so far could be described as follows: All participants, organisers and jurors stay at home and meet via video conference. Zoom  is a popular video conference tool which is well suited for presentations and - in our online GPT (see section \ref{sec:GPT} - for complete fight procedure (presentation, opposition, review). Opening meeting, introductory talks, and closing meeting including award ceremony could be done via video conference tool as well. Realising social events, for example coffee breaks, via online format are a little bit tricky, but virtual spaces like \emph{Wonder Me} \cite{didactics4} fit quite well. 

There could be a test run right before the actual tournament to make all people become familiar with the video conference tool that will be used. For our Online-GPT in december 2020 we did not do this, because GPT is a tournament for students at university, and all of our student participants use Zoom for listening lectures and taking part in seminars at university. The situation is different for tournaments addressed to students from school: A lot of them are not familiar with video conference tools used in tournaments. Hence in fact a test run is a good idea.

\subsection{Thoughts about pedagogic and didactic aspects of online tournaments}\label{sec:thoughts}
Research about didactic and pedagogic aspects of online tournaments is not yet available, and even research on the field of online (physics and science) teaching impacted by COVID-19 pandemic is still in its infancy (see e.g. \cite{didactics6}).

Nevertheless, we think that one of the most \emph{important didactic and pedagogic aspects} is \emph{fostering digital media literacy}\cite{digital}: Media literacy provides students (at university as well as students at school) with skills that will help them foremost think critically about media. It also cultivates other 21st-century skills like creativity, collaboration and communication, as well as increasing digital literacy skills through interacting with media, information and technology. Half of the participants of our Online-GPT 2020 are willing to become teachers at German grammar schools. They especially could not only use the their competencies in daily live, but pass them through the students at school while teaching.

Of course, there are also negative aspects coming along with online tournaments: It is the social component, that is missing, for example experimenting together or discussing in presence (students as well as jurors). 

\subsection{The time after COVID-19 pandemic}\label{sec:afterCOVID}
Logically one would expect that after an end of COVID-19 pandemic all tournaments will take place again in presence and at a venue. But - then all positive didactic and pedagogic aspects that are going along with online tournaments won't be present. 

So it could be a good concept for the future to come back to a tournament in presence, but to make preliminary rounds via online format.

When surveying participants of our Online-GPT, we came to the result that this could be a quite attractive idea, especially for those students who don't want to spent too much time (and money) with travelling to a (nice!) venue. 

That presence-and-online concept could be attractive for organisers as well, because organisation and realisation would be more easy and more cheep.

That epidemic has pushed education to become more aligned with modern technology and the future and we should keep the positive aspects for future tournaments, too.

\section{The Cartesian Diver}
\label{sec:divers}
\begin{figure}\centering
   \includegraphics[width=0.8\columnwidth]{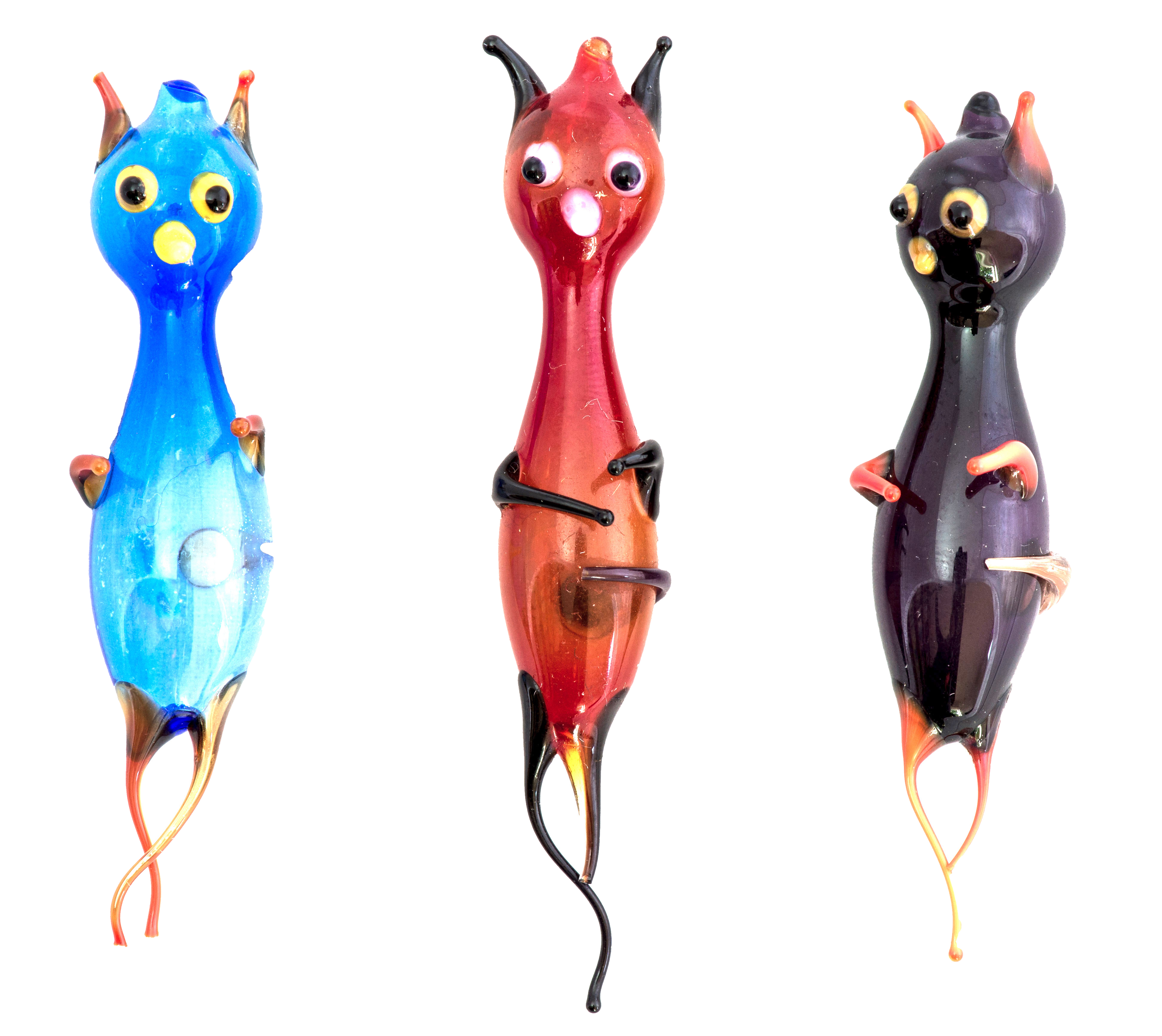}
    \caption{Commercially available Cartesian devils (or divers). The divers have an empty cavity inside the body and a tiny opening, e.g. on the top or at the end of the tail like the divers shown in the figure.}
\label{fig:cdevil} 
\end{figure}

The Cartesian diver, also known as Cartesian Devil, is an object that floats on water. It has a cavity and an opening where water can enter. Apart from this, the shape of the diver is arbitrary. The diver is placed into a container filled with water. Upon increasing the pressure in the container, the diver starts to sink. Figure \ref{fig:cdevil} shows a typical Cartesian devil.

\subsection{Theoretical introduction}\label{sec:theoryintro}
In this chapter, the principles that underlie the phenomenon of the sinking Cartesian diver are explained.

\paragraph{Archimedes' principle}

Archimedes' principle states that the upward buoyant force of an object in a liquid is equal to the weight of the liquid that the object displaces. It is the underlying principle for explaining why objects float or sink and will be used to derive necessary formulas.

Assume the diver with mass $m$ has initial cavity volume $V_i$, reduced cavity volume $V_0$ in starting position due to water (with a volume of $V_i-V_0$) entering the cavity and total volume $V_T$. Assume also that the diver floats on water at equilibrium in its starting position and has volume $V_{ext}$ above the water. We can apply Archimedes' principle with the buoyancy force $F_B$ and weight of the diver $F_G$ to find
\begin{eqnarray}
    &F_B=F_G  \nonumber \\
    \Leftrightarrow &m = (V_T-V_{ext}-(V_i-V_0))\rho ,\label{eq:V_ext}
\end{eqnarray}
where $\rho=\SI{997}{\kilogram\per\cubic\metre}$ is the density of water. We can observe that $V_{ext}$ depends only on the shape of the diver and its mass. 3D printing divers instead of using e.g. baking aroma tubes and trying to measure their volume is therefore an important improvement to this experiment in terms of uncertainties of measurement.

\paragraph{Pressure to sink a diver}

When applying pressure $p_1$ to the container, due to Pascal's law that pressure is applied through the incompressible liquid to the air inside the cavity of the diver. Assuming that the air can be approximated to behave like an ideal gas and that the thermodynamic process is isothermal, using Boyle's law we get
\begin{equation}\label{eq:pressure}
    p_0V_0=p_1V_1,
\end{equation}
where we use $p_0=\SI{101.325}{\pascal}$ the standard sea-level air pressure and \begin{equation}\label{eq:V_1}V_1=V_0-\Delta V\end{equation} is the volume of air inside the diver at pressure $p_1$. Assuming that the diver is, again, at equilibrium position at this pressure and completely sank below water, it displaces a volume of water equal to $V_T-\Delta V-(V_i-V_0)$ and we can use Archimedes' principle again:
\begin{equation}\label{eq:secondep}
    m=(V_T-\Delta V-(V_i-V_0))\rho.
\end{equation}

\begin{figure*}
\centering

 \centering
 \includegraphics[width=\columnwidth]{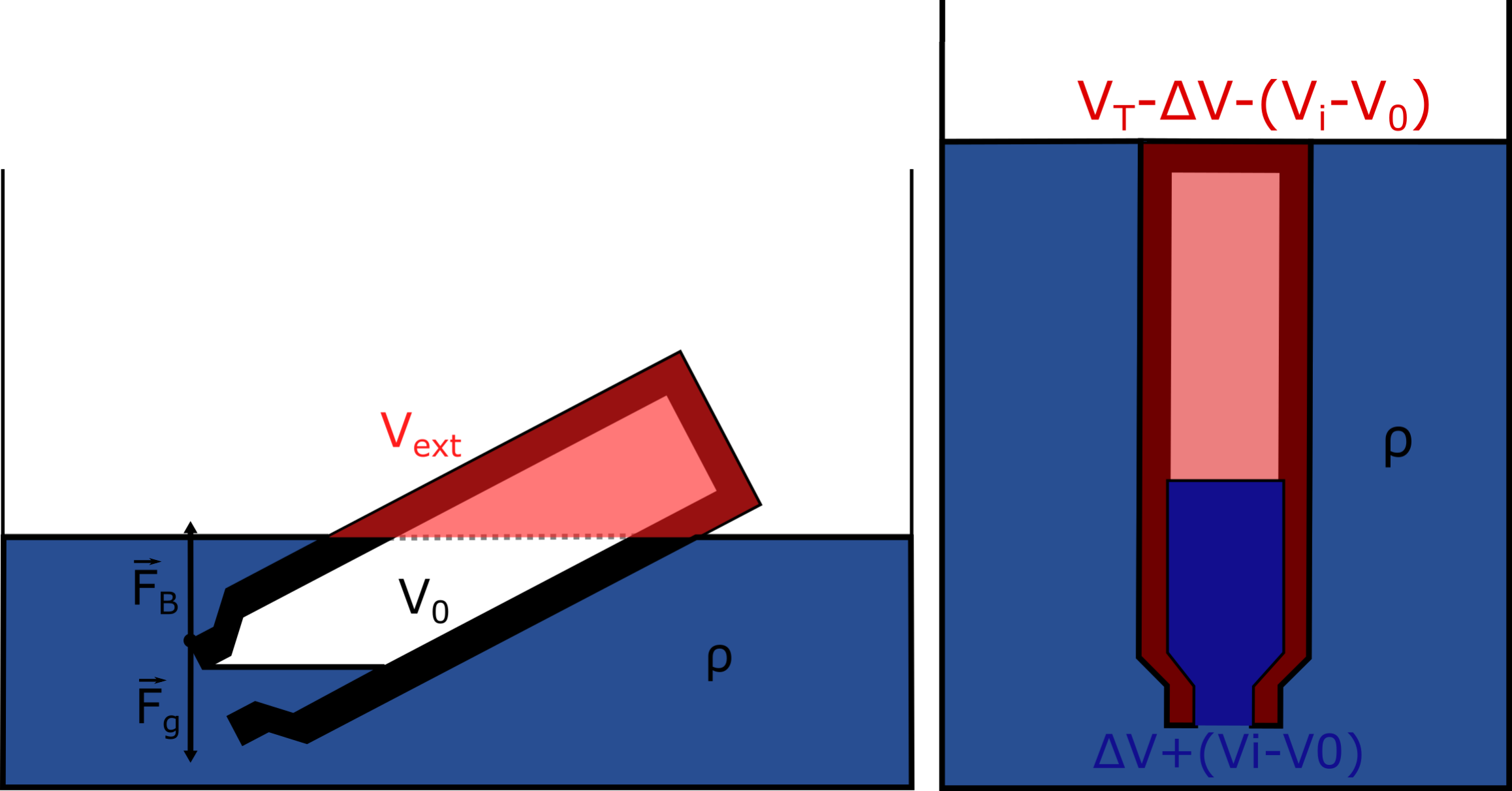}
    \caption{left: Volumes of divers at starting position (equation \ref{eq:V_ext}), right: Volumes of divers at fully submerged equilibrium position (equation \ref{eq:secondep}).}
\label{fig:DiverVolumes} 
\end{figure*}

The volumes of the divers at starting position (equation \ref{eq:V_ext}) and submerged equilibrium position (equation \ref{eq:secondep}) are illustrated in figure \ref{fig:DiverVolumes}.

Comparing equation \ref{eq:V_ext} and \ref{eq:secondep}, we find
\begin{equation}
    \Delta V = V_{ext}.
\end{equation}
Substituting this into equation \ref{eq:V_1} and the result into equation \ref{eq:pressure} and solving for $p_1$ we obtain
\begin{equation}\label{eq:pressurefinal}
    p_1=\frac{p_0}{1-\frac{V_{ext}}{V_0}}.
\end{equation}
We can therefore try to predict the pressure at which the divers sink using only their geometric values and finding $V_0$ by measuring $V_i-V_0$ by dropping the divers into water and measuring their weight $m_1$ (table \ref{tab:1}) after taking them out of the water and drying.
\subsection{Printing divers}

Cartesian divers can be prepared for experimentation in several ways. For example, small glass tubes (baking aroma), ink containers or straws can be used to create simple divers for use in schools. But in order to examine the relevant properties more closely, however, various parameters, such as mass and propulsive force, must be known and varied as exactly as possible. 
In recent years, the spread and use of 3D printers in schools and universities has increased. 3D printers enable the production of models and low-cost experimental materials in physics classes \cite{pusch}. In this case, 3D printing  enables enables an easy manufacturing of divers with varying geometric parameters, which are described in the next section. The divers were printed with the Fused Deposition Modeling process (FDM) which is most widespread in schools with the material type PETG. According to the data sheet \cite{datasheet}, this has a density of $\SI{1.27}{\gram\per\cubic\cm}$ as filament. We calculated that the density after the printout is $\SI{1.20\pm0.03}{\gram\per\cubic\cm}$, which may be due to the fact that individual air inclusions can arise during the printing process.

\subsection{Selection of parameters for the divers}
\begin{figure*}
\centering

 \centering
 \includegraphics[width=\columnwidth]{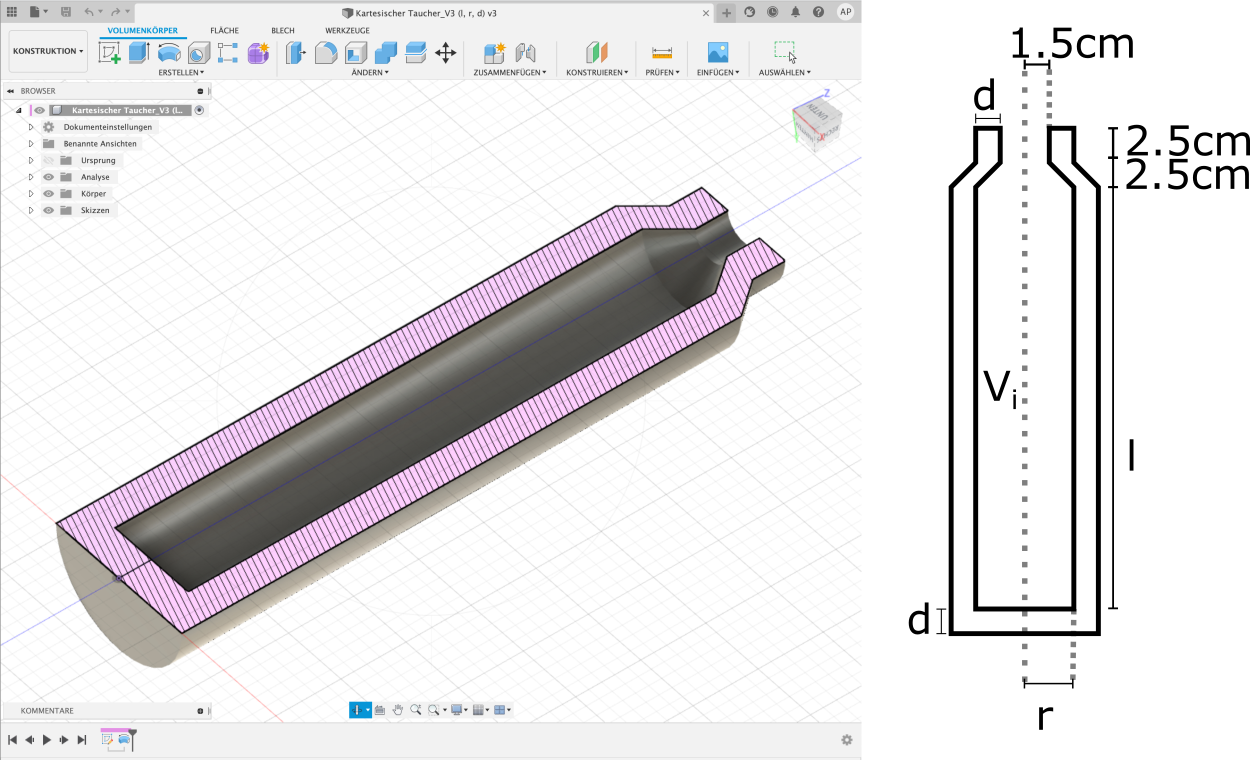}
    \caption{Design of divers in Autodesk Fusion 360 (left) and longitudinal section of the divers and parameters (right).}
\label{fusion} 
\end{figure*}

\begin{figure}\centering
   \includegraphics[width=0.6\columnwidth]{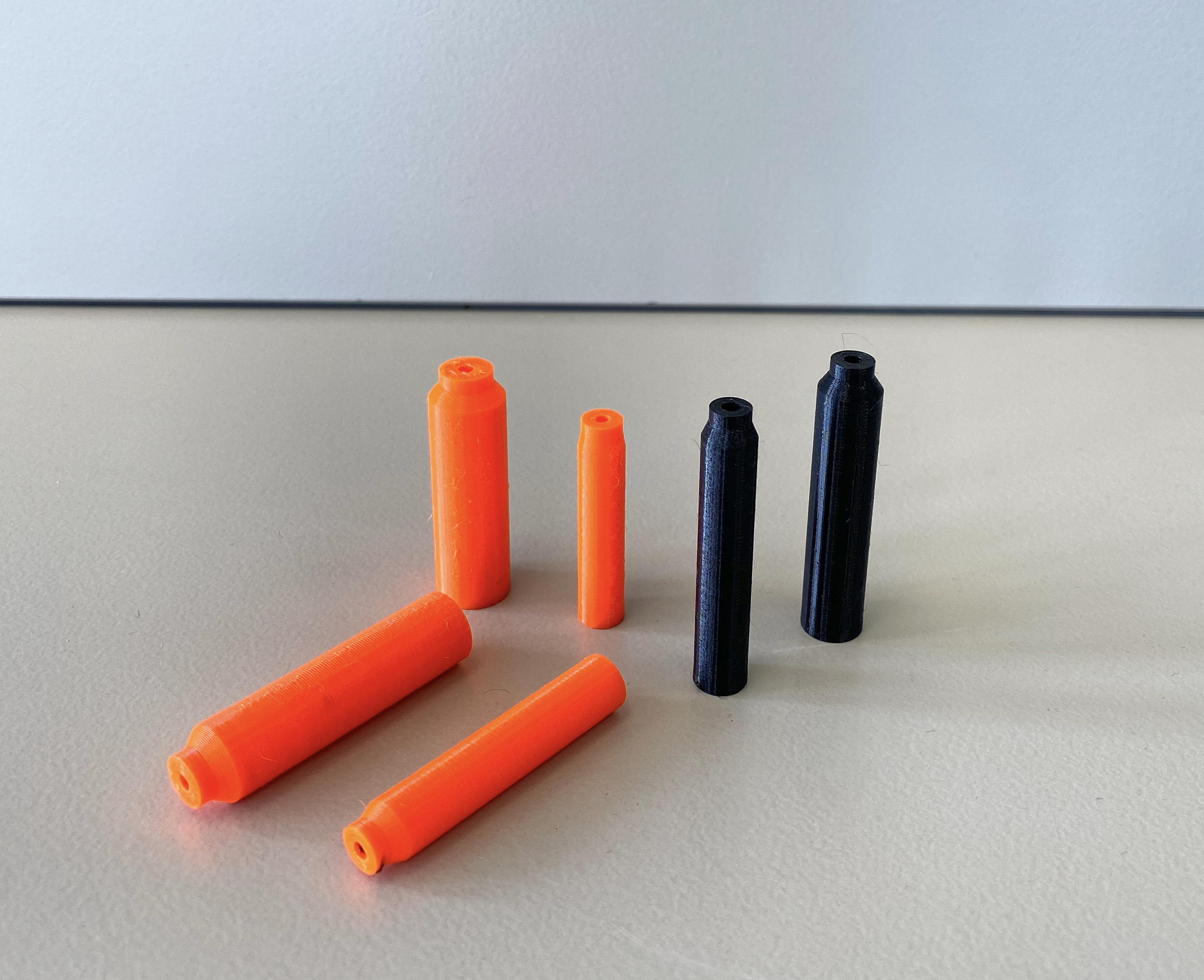}
    \caption{Printed Cartesian divers to investigate. Standing divers, left to right: diver no. 1, reverse diver, 4, 5. Lying, left to right: diver no. 3, 2.}
\label{fig:divers} 
\end{figure}

In order to vary the parameters buoyancy and mass, six divers were printed (figure \ref{fig:divers}) including one of them as the reverse cartesian diver (section \ref{sec:reverse}). The divers were parametrically constructed with Autodesk Fusion 360 (figure \ref{fusion}; \cite{pusch}). The parameters cavity radius $r$, wall thickness $d$ and length of main body $l$ were varied (table \ref{tab:1}) and $V_i-V_0$ was measured. Varying $l$ doesn't modify the value of $p_1$ significantly, but it is useful to distinguish the divers. Table \ref{tab:2} shows the calculated volumes and pressures at which the divers sink.

\begin{table}\centering
\caption{parameters of printed divers and mass, compare to figure \ref{fusion}}
\label{tab:1}
    \begin{tabular}{llllll}
        \noalign{\smallskip}\hline\noalign{\smallskip}
        diver no. & $r$  in \si{\milli\metre} & $d$  in \si{\milli\metre} & $l$  in \si{\milli\metre} & $m$ in \si{\gram} & $m_1$ in \si{\gram} \\
        \noalign{\smallskip}\hline\noalign{\smallskip}
        1 & $3.50\pm0.07$ & $3.00\pm0.07$ & $38.00\pm0.07$ & $5.021\pm0.004$ & $5.029\pm0.004$\\
        2 & $2.50\pm0.07$ & $2.00\pm0.07$ & $42.00\pm0.07$ & $2.569\pm0.004$ & $2.579\pm0.004$\\
        3 & $3.50\pm0.07$ & $2.50\pm0.07$ & $43.00\pm0.07$ & $4.540\pm0.004$ & $4.557\pm0.004$\\
        4 & $2.60\pm0.07$ & $1.80\pm0.07$ & $44.00\pm0.07$ & $2.249\pm0.004$ & $2.257\pm0.004$\\
        5 & $3.20\pm0.07$ & $2.10\pm0.07$ & $45.00\pm0.07$ & $3.210\pm0.004$ & $3.220\pm0.004$\\
        \noalign{\smallskip}\hline
    \end{tabular}
\end{table}

\begin{table}\centering
\caption{Inside cavity volume $V_0$, total volume $V_T$, external volume $V_{ext}$ and sinking pressure $p_1$ of all divers, see section \ref{sec:theoryintro}}
\label{tab:2}
    \begin{tabular}{lllll}
        \noalign{\smallskip}\hline\noalign{\smallskip}
        diver no. & $V_0$  in \si{\cubic\milli\metre} & $V_T$  in \si{\cubic\milli\metre} & $V_{ext}$  in \si{\cubic\milli\metre} &  $p_1$ in \si{\bar} \\
        \noalign{\smallskip}\hline\noalign{\smallskip}
        1 & $1520\pm60$ &  $5840\pm170$ & $790\pm170$ & $2.1\pm0.2$ \\
        2 & $860\pm50$ &  $3020\pm130$ & $440\pm130$ & $2.0\pm0.3$ \\
        3 & $1700\pm70$ &  $5470\pm170$ & $900\pm170$ & $2.1\pm0.2$ \\
        4 & $980\pm50$ &   $2990\pm130$ & $720\pm130$ & $3.8\pm0.6$ \\
        5 & $1500\pm60$ &  $4420\pm160$ & $1190\pm160$ & $4.8\pm0.6$  \\
        \noalign{\smallskip}\hline
    \end{tabular}
\end{table}

\begin{figure}\centering
   \includegraphics[width=0.4\columnwidth]{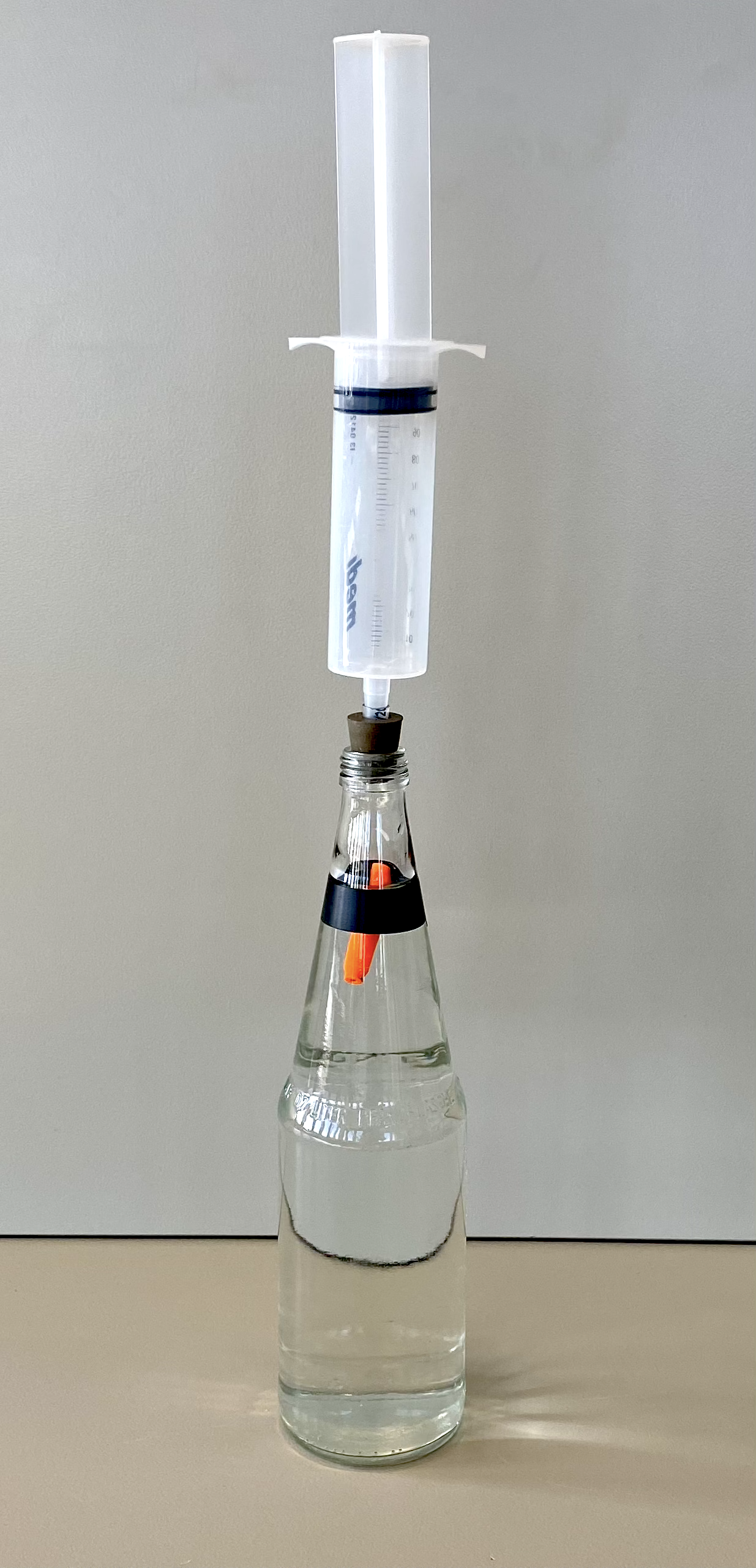}
    \caption{Experimental setup to investigate the pressure for diving by change of volume. The black tape marks the waterline that determines the required exact amount of air inside the experimental assembly of bottle and syringe. Also see the supplemental video 1 \cite{supplemental_1} for a video of the experiment.}
\label{fig:experiment} 
\end{figure}

\subsection{Experimental setup}
The diver is put into a filled bottle of water. With the experimental setup for the Cartesian diver, the ambient air pressure inside the bottle must be able to be increased. This can be done, for example, through a squeezable, sealed plastic bottle. To examine the model parameters, however, it is important to know the air pressure. This is possible with simple means, if the change in volume of the air is known, so that the air pressure can be approximated using the ideal gas law. For this we use a glass bottle to which a plastic syringe is connected. The change in volume can be read on the syringe. Figure \ref{fig:experiment} shows the experimental setup. 

With the initial total volume $V_T'$ of air inside the whole enclosed experimental setup and the change in volume $\Delta V'$ that is measured with the syringe, similarly to above with Boyle's law, we find
\begin{equation}
    p'_1=\frac{p_0}{1-\frac{\Delta V'}{V_T'}}.
\end{equation}
$V_T'$ consists of the air inside the connecting part of bottle and syringe, the syringe itself, the air inside the bottle and the air inside the diver at starting position.

\subsection{Results}

\begin{figure}\centering
   \includegraphics[width=0.6\columnwidth]{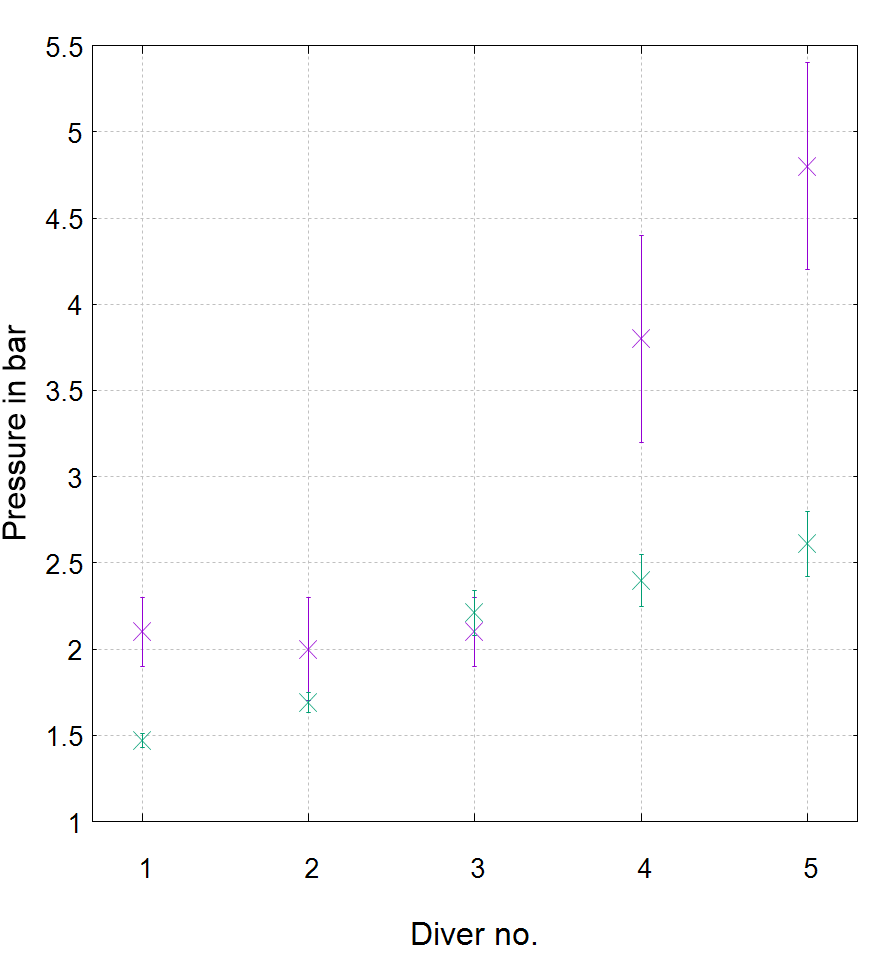}
    \caption{Pressures necessary to let the divers sink. Results from model calculations in purple and experimental results in green.}
\label{fig:results} 
\end{figure}

Table \ref{tab:3} and figure \ref{fig:results} show the results of the measurements. While the resuls for diver 2 and 3 lie within the uncertainty, the results for low and high pressures (diver no. 1 as well as 4 and 5) don't quite match up to the theoretical calculations. This is discussed further in section \ref{sec:discussion}.

\begin{table}\centering
\caption{Inside cavity volume $V_0$, total volume $V_T$, external volume $V_{ext}$ and sinking pressure $p_1$ of all divers, see section \ref{sec:theoryintro}}
\label{tab:3}
    \begin{tabular}{lllll}
        \noalign{\smallskip}\hline\noalign{\smallskip}
        diver no. & $p_1$ in \si{\bar} & $p'_1$ in \si{\bar}  \\
        \noalign{\smallskip}\hline\noalign{\smallskip}
        1 & $2.1\pm0.2$ & $1.47\pm0.04$ \\
        2 & $2.0\pm0.3$ & $1.69\pm0.06$ \\
        3 & $2.1\pm0.2$ & $2.21\pm0.13$ \\
        4 & $3.8\pm0.6$ & $2.40\pm0.15$ \\
        5 & $4.8\pm0.6$ & $2.61\pm0.19$ \\
        \noalign{\smallskip}\hline
    \end{tabular}
\end{table}

\subsection{Discussion}\label{sec:discussion}
The theoretical calculations don't quite match the experimental results, except for diver 2 and 3 where an intermediate pressure value is expected to sink the divers. Furthermore, the results from the model calculations have a higher relative uncertainty than the experimental results. This is due to the high uncertainty of the calculation of different volumes. Though there the 3D-printer posses a high accuracy, there still are inaccuracies in the 3D-printing process. A small change in the inside cavity volume (which will be affected by remaining air or water) can lead to big differences in the sinking pressure. Diver no. 3 has a smaller relative uncertainty of $p_1$ than diver no. 2 because it has higher inner radius $r$. Note also that due to the nature of equation \ref{eq:pressurefinal} inaccuracies strongly affect the cases where a higher pressure is predicted. This is because the closer $V_{ext}$ is to $V_T$, the closer the ratio $V_{ext}/V_0$ is to 1 and the higher the relative uncertainty of $1-V_{ext}/V_0$. In comparison, the experimental procedure of using a syringe to measure pressure is not as volatile to small inaccuracies and therefore more accurate in relation to the results. Note that the experiments have been performed at home, without laboratory equipment.

\subsection{The reverse Cartesian diver}\label{sec:reverse}

The reverse Cartesian diver is a diver that sinks to the bottom of the bottle normal pressure and rises to the top by lowering the pressure. It has to float on the water at first and only sinks due to water entering the cavity. Therefore, there is a sweet spot of the suitable parameters that has to be attained in construction of the diver. Table \ref{tab:4} shows the parameters of the reverse diver as well as the calculated sinking pressure and the second attached video shows the diver in action \cite{supplemental_2}. The fact that the diver is a reverse Cartesian diver shows that the diver's sinking pressure $p_1$ can not be below $\SI{1}{\bar}$, even though the standard deviation goes beyond this point. 

\begin{table}\centering
\caption{Parameters of the reverse cartesian diver}
\label{tab:4}
    \begin{tabular}{ll}
        \noalign{\smallskip}\hline\noalign{\smallskip}
        parameter & value \\
        \noalign{\smallskip}\hline\noalign{\smallskip}
        $r$ & $2.00\pm0.07$ \si{\milli\metre} \\
        $d$ & $2.00\pm0.07$ \si{\milli\metre} \\
        $l$ & $33\pm0.07$ \si{\milli\metre} \\
        $m$ & $1.868\pm0.004$ \si{\gram} \\
        $m_1$ & $1.877\pm0.004$ \si{\gram} \\
        $V_0$  & $450\pm30$ \si{\cubic\milli\metre} \\
        $V_T$  & $1970\pm90$ \si{\cubic\milli\metre} \\
        $V_{ext}$  & $83\pm9$ \si{\cubic\milli\metre} \\
        $p_1$ & $1.23\pm0.25$ \si{\bar} \\
        \noalign{\smallskip}\hline
    \end{tabular}
\end{table}

\section{Conclusions and Outlook}
\label{sec:conclusions}

The variable-controlled investigation of physical phenomena is an important part of learning and teaching physics in school and university. The use of 3D printing technology made it possible to easily vary the diver's parameters in order to investigate a connection between theoretical model and experiment. When dealing with the phenomenon, learners can vary the parameters which they have chosen themselves or implement other variants. Further variation and investigation possibilities are e.g. the variation of the size of the outlet opening or the implementation of a rotation around the axis in which the outlet occurs laterally. 

The Cartesian diver experiment shows another possibility of using 3D printers for home-made experiments. This experiment however requires high accuracy of the geometric values of small objects below the \si{mm} scale. In this scale FDM-3D-printers for domestic use come to their limits. Nevertheless, using a 3D printer made this experiment at all possible, especially the construction of the reverse Cartesian diver. 

The experimental way of measuring pressure using a syringe is easy to set up, low-cost and accurate enough for this experiment and can also have other applications. Our results indicate that the smallest inaccuracies in printing and therefore a good agreement between theory and experiments can be achieved for divers with a large inner radius (larger than \SI{30}{\milli\metre}). Furthermore, $V_{ext}$ and $V_0$ have to be sufficiently distinct.

Another possibility is the consideration of hydrostatic pressure and a fold catastrophe that happens when the object reaches depths where this pressure exceeds the pressure needed to sink the diver. When the diver reaches that point, it is bound to sink. A common glass bottle, however, is not deep enough to observe this phenomenon. Additionally, one can investigate the diver dynamics including friction in water. Both are discussed in \cite{fold}.

To conclude, by participating in Physics tournaments like the GPT, one has the opportunity of thoroughly dealing with Physics as a subject, but also with scientific methods and innovative technologies. In Physics education at university, students who are willing to become a teacher should also learn to convey a combination of factual knowledge in the subject and scientific methodology. Therefore, the GPT could be seen as a meaningful element with respect to learning and teaching didactics of physics.

\subsection{Acknowledgements}
We want to thank the members of the GPT network as well as everyone who helped to organize the Online-GPT 2020.
\newline

\end{document}